\documentclass{article}
\usepackage{smc2024}
\usepackage[caption=false, font=footnotesize]{subfig}
\usepackage{paralist}
\usepackage[figure,table]{hypcap}

\usepackage{csquotes}

\usepackage[whole]{bxcjkjatype}

\usepackage{alphabeta}
\usepackage{arabtex}
\usepackage[LFE,LAE,LGR,T2A,T1]{fontenc}
\usepackage[greek, russian, main=english]{babel}

\def\papertitle{Reconstructing the Charlie Parker Omnibook using an audio-to-score automatic transcription pipeline}

\author[]{\mbox{\firstname{Xavier}\lastname{Riley}\email{j.x.riley@qmul.ac.uk}}}
\author[]{\mbox{\firstname{Simon}\lastname{Dixon}\email{s.e.dixon@qmul.ac.uk}\orcid{0000-0002-6098-481X}}}

\affil[]{\department{Centre for Digital Music}\institution{Queen Mary University London}\country{UK}\affiliationtype{University}}

\completesetup

\title{\papertitle}
\begin{document}
	\capstartfalse
	\maketitle
	\capstarttrue

	\begin{abstract}
The Charlie Parker Omnibook is a cornerstone of jazz music education, described by pianist Ethan Iverson as ``the most important jazz education text ever published''. In this work we propose a new transcription pipeline and explore the extent to which state of the art music technology is able to reconstruct these scores directly from the audio without human intervention. Our pipeline includes: a newly trained source separation model for saxophone, a new MIDI transcription model for solo saxophone and an adaptation of an existing MIDI-to-score method for monophonic instruments.

To assess this pipeline we also provide an enhanced dataset of Charlie Parker transcriptions as score-audio pairs with accurate MIDI alignments and downbeat annotations. This represents a challenging new benchmark for automatic audio-to-score transcription that we hope will advance research into areas beyond transcribing audio-to-MIDI alone.

Together, these form another step towards producing scores that musicians can use directly, without the need for onerous corrections or revisions. To facilitate future research, all model checkpoints and data are made available to download along with code for the transcription pipeline. Improvements in our modular pipeline could one day make the automatic transcription of complex jazz solos a routine possibility, thereby enriching the resources available for music education and preservation.
	\end{abstract}

	\section{Introduction}\label{sec:introduction}

The automatic transcription of music from audio to a readable score represents a challenging area of research within the field of music information retrieval (MIR). Success in this task has the potential to revolutionise the way musicians learn, teach, and preserve music. Despite its significance, the task of converting an audio recording to a score of similar quality to those of professional engravers remains a challenging and open research area.

Among the many genres of music, jazz stands out as particularly complex due to its improvisation, syncopation, and swing-feel elements that are quintessential yet difficult to quantify and transcribe. The work of Charlie Parker, a leading figure in the development of bebop and modern jazz, demonstrates many of these characteristics. Parker's solos are not only technically demanding but also helped to define styles and idioms that have been adopted by subsequent generations of jazz musicians. This helps to explain why the Charlie Parker Omnibook\cite{omnibook} has been in print continuously since its publication in 1978. This collection of 60 transcriptions of his solos is a widely used resource for jazz students and professionals alike.

This paper investigates a pipeline approach to automatic music transcription (AMT) that leverages recent advances in source separation, MIDI transcription, and score generation technologies. By focusing on the solos of Charlie Parker, we aim not only to contribute a new tool to the field of music education but also to address a gap in the literature concerning the automatic transcription of complex jazz solos from audio to score. The choice of Charlie Parker as the subject of this study is deliberate: his work presents a challenging benchmark for evaluating the efficacy of AMT systems on material that is culturally significant and still widely studied today.


\setcounter{bottomnumber}{0}

\section{Related work}\label{sec:related_work}

The transcription pipeline we propose in this work incorporates several topics, each of which could be discussed in depth. Due to length constraints we give an overview of important work and refer to more detailed summaries elsewhere where necessary.

\subsection{Jazz transcriptions}

Transcription has been an essential part of jazz education and musicology as jazz is mainly an aural tradition with few written sources. Regarding Charlie Parker specifically, an early example of academic work was Thomas Owens's (1974) PhD thesis \cite{owens1974}, in which he transcribed a large number of Parker's solos leading to the identification of approximately one hundred melodic formulas. These range from concise four-note motifs to expansive multi-measure phrases. Even today, this extensive analysis provides inspiration for what automated approaches may enable.

Shortly afterward, the Charlie Parker Omnibook was published in 1978 and has been in print ever since. This contains 60 solos transcribed by trumpeter Ken Sloane with assistance from saxophonist Jamey Aebersold.

In the age of computational analysis, digital formats and scores are required. The most ambitious work to date is the Weimar Jazz Database\cite{pfleiderer2017inside}\footnote{\url{https://jazzomat.hfm-weimar.de/dbformat/dboverview.html}} which encompasses approximately 200,000 tone events from 456 monophonic solos by 78 jazz masters, providing high-quality score and MIDI transcriptions. These transcriptions were carried out with a semi-automated pipeline involving manual corrections. The dataset includes 12 works by Charlie Parker.

Interest in generative modelling of Parker's improvisations led to the release of a digitised version of the Omnibook \cite{deguernel}. 50 of the 60 scores were released in MusicXML format to aid further research.

More recently, the FiloSax dataset by Foster and Dixon \cite{filosax} includes 240 full length scores of jazz saxophone performances along with audio stems and copious metadata. These resources are key to enabling the transcription pipeline which we describe later in this work.


\subsection{Source separation}

Music source separation has seen a huge increase in interest with the advent of capable neural network models. The objective is to extract an individual part (stem) from an audio mixture by learning a source specific filter. This is a large and active research area and we refer elsewhere \cite{sounddemixingchallenge} for a recent survey and discussion of state-of-the-art approaches.

For this work we use Demucs\cite{demucs}, a widely used solution proposed by researchers at Facebook AI. We note that separation into drums/bass/vocals/other is widely available due to the MDB18 dataset\cite{mdb18}, however to date there are no freely available models for separation of saxophone. This work addresses this gap by training and releasing a solo saxophone separation model which separates into ``sax'' and ``other'' stems. Details of the training procedure are discussed in section \ref{subsec:transcription_pipeline}.


\subsection{Transcription to MIDI}

Automatic music transcription generally aims to convert musical audio into a useful symbolic representation such as MIDI. 
Benetos et al.\ \cite{benetos_amt} claim that monophonic transcription is a solved problem, however recent work \cite{crepe_notes} suggests that this is not true for more intricate sources like jazz saxophone and traditional Irish flute.

Historically, approaches to the monophonic transcription task have typically involved framewise pitch estimation followed by some method to segment the pitch contour into discrete notes. Pyin \cite{pyin} (in particular the Pyin notes function) is still a popular tool for this purpose, but its accuracy when classifying at the note level is not sufficient to avoid manual corrections. CREPE Notes\cite{crepe_notes} improves on this but it is still limited to 82.31\% note level F-measure on the FiloSax dataset.

Higher levels of accuracy have been achieved on piano recordings in recent years due to high quality datasets and several capable machine learning architectures for transcription \cite{kong}. The issue of instruments beyond the piano was recently addressed\cite{maman, hiresguitar} by using scores and existing transcriptions which were realigned to audio to produce training data.


\subsection{Score layout}

The process of converting a human performance to a musical score is typically performed by highly skilled human transcribers. Computational approaches have had limited success so far in this task. Performance timing must be interpreted in terms of musical units on a metrical grid and note durations must be chosen which allow the score to be parsed with minimal difficulty by human performers. 

The qparse library\cite{qparse} handles monophonic score layout by combining these two stages, using a probabilistic grammar defined by the user. We adopt this method in our work, training a new grammar based on Charlie Parker's notated scores, as described in section \ref{subsec:transcription_pipeline}.

Quantising a note to its nearest position on the grid is an obvious approach, however this does not always account for rhythmic variation in expressive human performance such as the use of swing feel and playing ``behind the beat''. We explore these issues further in section \ref{sec:discussion}.
Polyphonic score layout is more challenging and beyond the scope of this work and we refer the reader to \cite{lele} for more information.


\subsection{Other pipelines and alternative approaches}

The separate-transcribe-notate pipeline that we propose has been used before, however not in a fully automated setting. Work on transcribing jazz bass\cite{filobass} used a similar approach but the result required a number of manual corrections.

Large scale jazz transcription has similarly relied on automated approaches with corrections by human annotators\cite{pfleiderer2017inside,guitarset}. The Dig that Lick project\cite{dtl} attempted fully automatic extraction of MIDI data for jazz solos but did not include the conversion to musical score. Their approach \cite{basaran} extracted main melody contours directly from audio mixtures which were then segmented into notes. Melodia\cite{melodia} was the previous state of the art for melody estimation from monophonic and polyphonic sources. An alternative approach \cite{jointist,cerberus} combines the source separation and transcription stages as a joint learning problem, but these techniques do not produce score output.

In the latter stages of completing this work, we became aware of Martínez et al.\ \cite{martinez}, who present an approach for end-to-end audio-to-score (A2S) transcription, specifically applied to a newly constructed corpus of real and synthetic saxophone recordings. The corpus is smaller than FiloSax (3 hours vs.\ over 20 hours) and contains simpler melodic material such as musical scales and short phrases. Their method attempted to predict a text-based score representation (Kern) directly as model output, which differs from the pipeline approach we have taken in this work. They obtained similar results for their synthetic and real world audio data which suggests that the approach is viable, however the model weights were not available at the time of writing so we were unable to perform a direct comparison. This will be a subject of future work. 

In our work, we chose to investigate a more modular pipeline, so that any improvements in performance of system components will directly benefit our pipeline. 


\section{Method}\label{sec:method}

\subsection{Dataset production}\label{subsec:dataset_production}

We used an existing corpus of digital scores from the Omnibook \cite{deguernel}. This consists of 50 of the 60 original tracks in the published Omnibook, digitised to MusicXML. We hope to digitise the remaining 10 tracks as part of future work. We sourced recordings that corresponded to each transcribed segment and uploaded these to the SoundSlice platform\footnote{\url{https://www.soundslice.com}}. The first author then added downbeats manually to all 50 tracks, using the variable speed and fine adjustment settings that the platform offers. Only sections corresponding to the notated scores were annotated with downbeats. This typically includes the main theme and the saxophone solo. In most cases these form a continuous segment but additional metadata is included to account for gaps where other solos occur.

Given the age of the source material (recorded in live and studio settings before Parker's death in 1955) there were a number of issues with performances not being reproduced at concert pitch (A440Hz). We first attempted to correct for this using an automatic approach\footnote{\url{https://librosa.org/doc/main/generated/librosa.estimate_tuning.html\#librosa.estimate_tuning}} however on review these were found to be inaccurate, potentially due to the poor recording quality. We chose to use the SoundSlice pitch adjustment feature to make manual global pitch adjustments. These adjustments were applied to the source audio files using the rubberband library\footnote{\url{https://breakfastquay.com/rubberband/}}.

With downbeats and intonation resolved, we then aligned the human transcriptions (in MIDI representation) to the audio of the solo saxophone stem. This was done using Dynamic Time Warping (DTW, as implemented in the SyncToolbox \cite{synctoolbox}). To fine-tune the DTW path, the MIDI note onsets were aligned to the frame level activations of a transcription model using a high-resolution alignment procedure \cite{hiresguitar}. 


To aid further research, we make the dataset freely available to other researchers. This includes MusicXML scores, downbeats, performance-aligned MIDI files, links to the original recordings and extracted saxophone stems tuned to A440Hz standard. These can be accessed via the supplementary site\footnote{\url{https://aim-qmul.github.io/SaxTranscriptionPipeline/}}.


\subsection{Transcription Pipeline}\label{subsec:transcription_pipeline}

We now describe the proposed transcription pipeline for the task of transcribing a score from an audio recording containing saxophone.

The first stage is beat estimation, for which we chose Madmom \cite{madmom} for its strong performance and convenient implementation. This estimates the beat, downbeat and time signature from the original audio mix which are later used in score estimation. Initially we found that the default parameters led to a number of failures on the Omnibook recordings. To address this we first obtain a rough estimate of global tempo from the transcribed MIDI using the distribution of inter-onset intervals \cite{dixon2001automatic}. We also constrain the minimum tempo to 15 BPM below the prior estimate (to avoid picking a tempo which is too slow) and the maximum to 350 BPM (to account for the extreme tempi found in bebop music). Empirically this was found to improve tempo estimation for this dataset.

The next stage is to separate the saxophone from the audio mixture. The Demucs source separation model (version 3) \cite{demucs} was trained using audio from the FiloSax dataset \cite{filosax}. The complete set of 240 tracks was used for training to maximise the amount of available training data, with tracks divided into 6 second segments. The model was trained for 180 epochs with a learning rate of $0.001$ and a batch size of 8.

For transcribing the separated saxophone stem to MIDI, we train a new model based on the high resolution piano transcription work by Kong et al.\ \cite{kong}. We take an 80-10-10\% split (by pieces) of the FiloSax dataset. Each recording is divided into 10-second
samples, with a hop size of 1 second. We used a learning rate of $10^{-4}$ and a batch size of $8$. The training data came from tenor saxophone and so audio was augmented with random pitch shifts, continuously distributed between \(\pm 5\) semitones. This helps to match the registers of the alto and baritone saxophones respectively and increases training variety given the relatively small dataset. Training ran for 27.5K steps, approximately 55 epochs,
scaling the learning rate by 0.9 every 5K steps.

Finally for the MIDI-to-score stage we used the ``qparse'' package\cite{qparse}. This handles rhythmic quantisation of MIDI elements along with score engraving by treating it as a parsing problem using a user-specified grammar which defines probabilistic transition states between note types. We first defined a grammar based on the transitions observed in the Omnibook data using the method in \cite{qparsemodeling}, however this led to an overly complex grammar which led us to simplify some of the rules empirically. This was done to reduce the frequency of triplets in favour of quavers in the predicted scores. The final grammar definition is available with the code that is released alongside this work via the supplementary site.


\section{Results}

\subsection{Beat tracking}
\label{sec:results_beat_tracking}

We include results for automatically estimated downbeats in Table \ref{tab:downbeat_results}. The method used is described in section \ref{subsec:transcription_pipeline}. From the results we see that the automatic estimation is generally poor, albeit with a wide variance. As all the pieces in the dataset are in 4/4, we also include results for the optimal downbeat placement within a bar (i.e.\ selecting the best of all possible ``rotations''). The improvement when rotations are included suggests that the beat tracking is relatively accurate, but that downbeat estimation is challenging on this source material. Since it requires little user effort to specify the correct rotation, it may be desirable to compute multiple transcriptions for each possible downbeat rotation and allow the end-user to select the best result.

\begin{table}
 \begin{center}
 \begin{tabular}{|l|l|l|}
  \hline
  Method (Madmom) & Mean F-measure & Max \\
  \hline
  w/ init bpm & $29.20 \pm 25.11$ & 92.52 \\
  w/ init bpm and rotations & $61.27 \pm 30.57$ & 97.74\\
  \hline
 \end{tabular}
\end{center}
 \caption{Downbeat F-measure results calculated by \textit{mir\_eval} for the proposed Charlie Parker dataset. The values are expressed as percentages and a default tolerance of 70ms was used.}
 \label{tab:downbeat_results}
\end{table}

\subsection{Solo saxophone source separation}\label{sec:results_source_sep}

\begin{table}
 \begin{center}
 \begin{tabular}{|l|l|}
  \hline
  Method & SDR (dB) \\
  \hline
  LALAL.AI Wind Stem & $14.03 \pm 2.38$ \\
  Ours (Demucs + FiloSax) & $\textbf{14.22} \pm 2.10$ \\
  \hline
 \end{tabular}
\end{center}
 \caption{Source Separation: Signal-to-Distortion Ratio (SDR) results calculated by \textit{mir\_eval} for two methods. The SDR values are expressed in decibels (dB) and include the standard deviation. Higher is better.}
 \label{tab:sdr_results}
\end{table}

We evaluate our source separation model on a private dataset of 10 professionally produced jazz quartet recordings featuring saxophone. To the best of our knowledge, there are no recent source separation models for saxophone discussed in the literature, therefore we have compared our results with a commercial offering\footnote{\url{https://www.lalal.ai/}} to provide a baseline. As we can see in Table \ref{tab:sdr_results}, our method performs better, but it should be noted that the Lalal.ai results were for a more general ``Wind'' stem, whereas ours was specifically trained for saxophone.
    
    
\subsection{MIDI transcription}\label{ssec:results_midi}

This stage of the pipeline was evaluated on two datasets using standard metrics for note-level accuracy (no offsets, 50ms tolerance) provided by the mir\_eval library. Evaluation results for a test split of the FiloSax dataset are shown in Table \ref{tab:fs_test_midi}, and results for the Charlie Parker Omnibook dataset are shown in Table \ref{tab:cp_midi}.

Here we estimate a MIDI transcription using the source separated saxophone stems as input, which is then compared with the aligned MIDI created from the ground truth scores. We evaluate our method alongside CREPE Notes, which previously demonstrated state-of-the-art results on FiloSax over a number of other methods \cite{crepe_notes}. We also include results for Ba\c{s}aran et al.\cite{basaran}, as this was the method selected to produce automated transcriptions for the Dig That Lick project.

On the FiloSax data, our method demonstrates a large improvement over previous methods, but this should be taken with caution as the test data is from the same distribution as the training set. On the out-of-distribution Omnibook data, 
our method also achieves the highest F-measure score, but the results are considerably lower than those achieved for FiloSax. The three systems tested gave results that were respectively over 12\%, 16\% and 20\% worse than the Filosax results, suggesting that the Omnibook can be considered a very challenging dataset.

\begin{table}
 \begin{center}
 \begin{tabular}{|l|l|l|l|}
  \hline
  & $P_{50}$ & $R_{50}$ & $F_{50}$ \\
  \hline
  CREPE Notes \cite{crepe_notes} & 82.35 & 83.05 & 82.67 \\
  Ba\c{s}aran et al.~\cite{basaran} & 87.68 & 92.57 & 90.01 \\
  Ours & \textbf{96.47} & \textbf{95.90} & \textbf{96.19} \\
  \hline
  
 \end{tabular}
\end{center}
 \caption{Note level transcription (FiloSax): Results for note-level transcription accuracy (no offsets) on a 25 track test split of FiloSax. $P_{50}$, $R_{50}$, and $F_{50}$ are Onset-only Precision, Recall and F1-measure, expressed as percentages, at 50ms resolution.}
 \label{tab:fs_test_midi}
\end{table}

\begin{table}
 \begin{center}
 \begin{tabular}{|l|l|l|l|}
  \hline
  & $P_{50}$ & $R_{50}$ & $F_{50}$ \\
  \hline
  CREPE Notes \cite{crepe_notes} & 70.93 & 70.63 & 70.41 \\
  Ba\c{s}aran et al.~\cite{basaran} & 70.60 & \textbf{77.98} & 73.68 \\
  Ours & \textbf{74.70} & 76.96 & \textbf{75.43} \\
  \hline
  
 \end{tabular}
\end{center}
 \caption{Note level transcription (Omnibook): Results for note-level transcription accuracy (no offsets) on the proposed Omnibook dataset alignments using stems from section \ref{sec:results_source_sep}. Abbreviations are described in Table \ref{tab:fs_test_midi}.}
 \label{tab:cp_midi}
\end{table}


\begin{figure*}
    \centering
    \includegraphics[width=\textwidth,trim={0 15cm 0 0cm},clip]{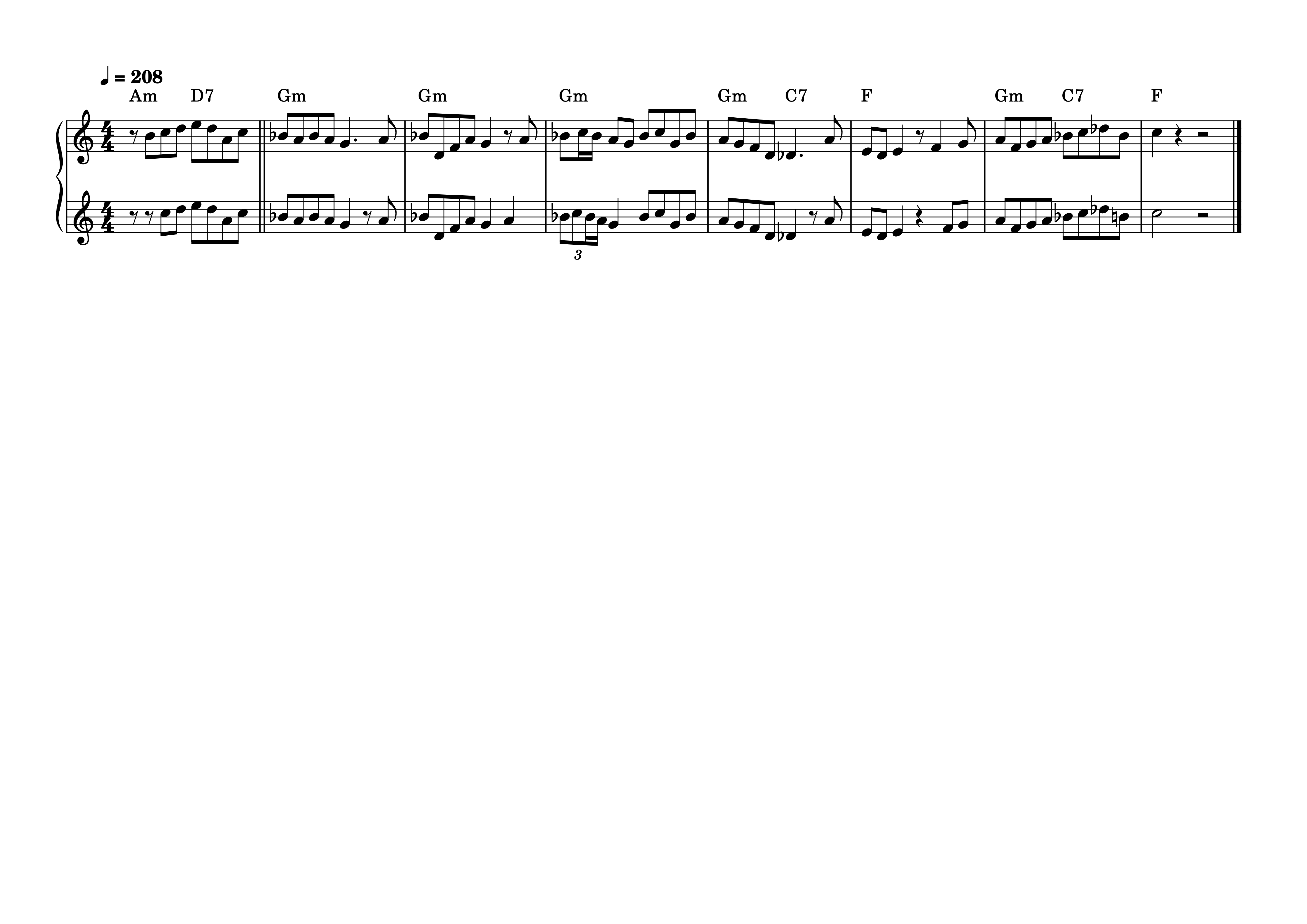}
    \caption{An extract of ``Marmaduke'' with ground truth on the upper staff and our transcription on the lower staff.}
    \label{fig:score_example1}
\end{figure*}

\begin{figure*}
    \centering
    \includegraphics[width=\textwidth,trim={0 15cm 0 0cm},clip]{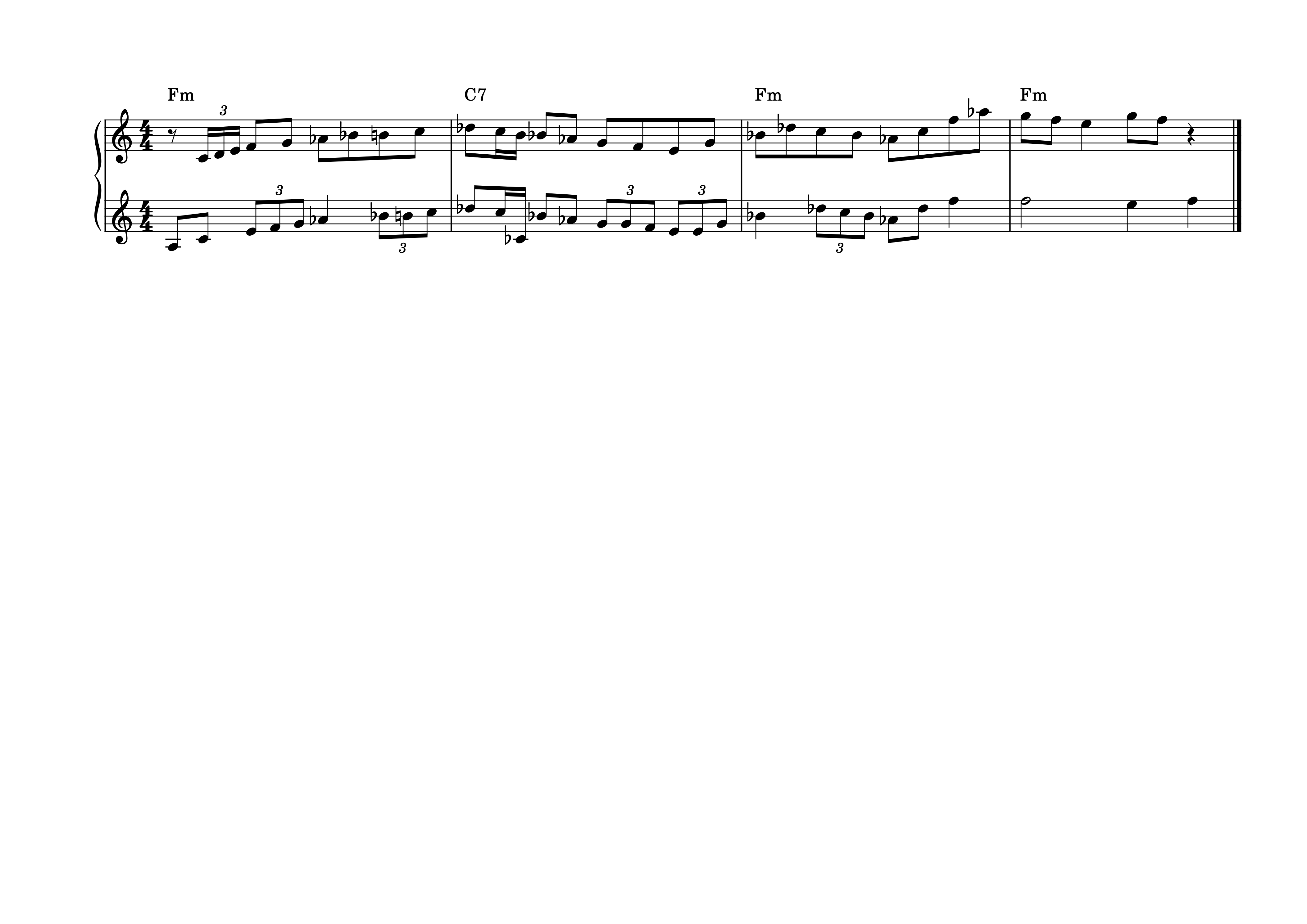}
    \caption{An extract of ``The Bird'' with ground truth on the upper staff and our transcription on the lower staff.}
    \label{fig:score_example2}
\end{figure*}

\subsection{Score timing}\label{ssec:score_layout}

The task of producing a score from a MIDI representation of a human performance has received less attention than the audio-to-MIDI stage and as a consequence there is less agreement in the literature on which metrics are most appropriate. We have used a subset of the metrics proposed by Cogliati et al.~\cite{cogliati} to give an indication of the edit distance between our transcribed scores and the ground truth.

For comparison we include a baseline from the Music21 package \cite{music21}, which is used to import the transcribed MIDI with \texttt{quantizePost=True}. This is to give an indication of score quality when handled by basic quantisation alone.

The results in Table \ref{tab:score_edit} show mean error rates, normalised by the number of notes in each piece. We find that our proposed qparse workflow requires fewer edits than the basic quantisation method, showing particular efficiency regarding the number of rests that need to be edited (34.50\% insertions vs 791.93\% for the baseline method). This means our method requires less work to reach the level of a human transcribed score, however there is still room for improvement. 

\begin{table}
\begin{center}
\resizebox{\columnwidth}{!}{
\begin{tabular}{|l|cc|cc|l|}
\hline
& \multicolumn{2}{c|}{Note} & \multicolumn{2}{c|}{Rest} & \\
\cline{2-5}
& Insert & Del & Insert & Del & \multicolumn{1}{c|}{TimeSig} \\
\hline
qparse \cite{qparse} & 23.26 & 54.74 & 34.50 & 80.23 & 2.55 \\
baseline \cite{music21} & 45.36 & 48.41 & 791.93 & 41.26 & 2.69 \\
\hline
\end{tabular}
}
\end{center}
\caption{Score quantization: Mean error rates for score timing expressed as percentages of the ground truth score length. Results are shown for the 31 scores that qparse could successfully process. The baseline is described in section \ref{ssec:score_layout}.}
\label{tab:score_edit}
\end{table}

We also present some excerpts of transcribed scores with their associated ground truth in Figure \ref{fig:score_example1} and Figure \ref{fig:score_example2}. The complete set of scores is available for detailed comparison via the supplementary site\footnote{\url{https://aim-qmul.github.io/SaxTranscriptionPipeline/}}. From these we see that our proposed method performs well at producing ``readable'' scores, at the expense of some rhythmic accuracy. Offbeat eighth notes are often quantised to be on the beat, for example, which suggests that further work is required to refine the grammar rules provided to qparse. In Figure \ref{fig:score_example2} we also see issues with tracking pitches in the upper register of the alto, as discussed in Section \ref{sec:discussion} below.

\subsubsection{Stability of qparse}\label{sssec:qparse_stability}

In this work we adopted the latest version of qparse, however we encountered a number of issues that remain unresolved at the time of writing. This resulted in only 31 out of 50 scores being available for comparison, with the others failing to find a valid parsing result. In the released code we provide a fallback option to offer more basic parsing in cases where qparse fails, and we will continue to work with qparse authors to improve the robustness of the code.

\section{Discussion}\label{sec:discussion}

In attempting to reconstruct this corpus of saxophone transcriptions, we have encountered several challenges described in the prior sections. In spite of these, we are pleased to note that our proposed pipeline obtains the highest accuracy of existing methods for the audio-to-MIDI stages and that the MIDI-to-score stage outputs useful scores in many cases.

Some of the difficulties we faced in producing scores from performance MIDI were surprising, even in the relatively simple monophonic case which our work focused on. We hope to explore this issue further in upcoming work to find a robust method for conversion from performance timing to score timing. The goal would be to handle the trade off between rhythmic accuracy and score complexity. We also consider that renewed research into metrics and measures of score transcription quality would be beneficial in this area.

A particularly challenging problem was that of beat detection, which traditionally assumes that the performers share a concept of a single shared pulse. In jazz however, there has been some work to suggest that participatory discrepancies (PDs) are a central feature of the music. This is where a soloist, for example, might ``lay back'' and phrase their part deliberately late\cite{butterfield}. This idea is obliquely supported in a 1949 DownBeat interview with Charlie Parker himself: when pushed to define the new genre of ``bop'' he gave the answer:

\begin{displayquote}
``The beat in a bop band is with the music, against it, behind it,'' Charlie said. ``It pushes it. It helps it.''
\end{displayquote}

This suggests that his placement of notes (``the music'') was deliberately manipulated in reference to the pulse (``the beat'') and indeed, this is something we observe in our data. For example, we see that short note values at the ends of bars are often quantised to the start of the next bar. Given the lack of synchrony between performers, it is clear that future MIDI-to-score solutions need to be flexible with the concept of a downbeat in order to succeed in transcribing jazz in a human-like fashion.

The separation and MIDI transcription parts of our pipeline are shown to perform well, however we feel that their generalisation to all types of saxophone could be enhanced further. Empirically, our system struggles with the extreme upper register of the alto saxophone (altissimo, harmonics etc.) but we believe this is due to the available training data being in a conservative range of the tenor saxophone. This could be improved in future iterations with access to more diverse data (including alto saxophone) and with more aggressive data augmentation strategies.

Despite these challenges, we do find that the best case outputs of our proposed method produce usable scores. With further refinement of this pipeline we take a step towards transcribing music accurately at scale. This opens up new possibilities for management of large collections, and also the potential for time savings for the transcription workflows of individual musicians.

	
\section{Conclusions}

In this work we present a new pipeline for end-to-end transcription of solo saxophone from audio to score. Each stage in the pipeline is discussed and evaluated where possible, with state-of-the-art results for source separation of saxophone and MIDI transcription of solo saxophone.

Alongside the transcription pipeline, we release a new dataset of audio-score pairs of 50 Charlie Parker recordings. These feature accurate alignments and timing information which allow for use in tasks such as transcription, expressive performance analysis and generative modelling.

We also publish our automated transcriptions alongside the ground truth scores to demonstrate what is currently possible in difficult cases of monophonic transcription from audio to score. The results are not yet at the level of a human expert, but are a promising step towards more reliable transcription methods in future. We hope these will serve as a baseline to stimulate future research in this area.

	
	\begin{acknowledgments}
The first author is a research student at the UKRI Centre for Doctoral Training in Artificial Intelligence and Music, supported by UK Research and Innovation [grant number EP/S022694/1]. We would like to thank Florent Jacquemard and Lydia Rodriguez de la Nava for their co-operation and assistance with using qparse for this work.
	\end{acknowledgments} 
	
	\bibliography{smc2024}
	
\end{document}